\documentclass[aps,prb,twocolumn,showpacs,amsmath,floatfix]{revtex4}
\usepackage{graphicx}
\usepackage{dcolumn}

\begin{document}
 \topmargin-0.8cm


\title{
Order-N implementation of exact exchange in extended systems}

\author{Xifan Wu, Annabella Selloni, and Roberto Car}

\address{Chemistry Department, Princeton University, Princeton, NJ
08544-0001,USA}

\date{\today}

\begin{abstract}

Exact (Hartree Fock) exchange is needed to overcome some of the limitations of local 
and semilocal approximations of density functional theory (DFT).
So far, however, computational cost has limited the 
use of exact exchange in plane wave calculations for extended systems. We show that 
this difficulty can be overcome by performing a unitary transformation from Bloch to 
Maximally Localized Wannier functions in combination with an efficient technique to 
compute real space Coulomb integrals. The resulting scheme scales linearly with system 
size. We validate the scheme with representative applications.

\end{abstract}


\maketitle


\marginparwidth 2.7in

\marginparsep 0.5in

\def\xwm#1{\marginpar{\small XW: #1}}
\def\rcm#1{\marginpar{\small RC: #1}}
\def\asm#1{\marginpar{\small AS: #1}}

Electronic structure calculations based on density functional theory (DFT)  have been very successful in studies of molecular and condensed matter systems. To date most DFT applications to extended material systems have used the local density approximation (LDA)  or the semi-local generalized gradient approximation (GGA)  for exchange and correlation \cite{DFT}. 
These approximations are numerically efficient but suffer from serious drawbacks. 
In particular, the spurious self-interaction of each electron with itself, 
occurring with local and semi-local functionals, may lead to a poor 
description of tightly bound electronic states \cite{Yang}. 

These deficiencies are less severe when hybrid functional approximations for exchange and correlation are adopted.\cite{Hybrid,Kummel}  In this approach some exact exchange energy is mixed into the exchange-correlation energy functional. Extensive applications to molecular systems have shown that hybrid functionals are generally superior to GGA in the description of structural and electronic properties \cite{Kummel}.  Applications to extended systems have been so far relatively scarce, even though available studies suggest that hybrid functionals should provide a better description of the electronic properties of insulating materials.\cite{Hybrid_solid_chemistry, localized_state, PBE0_vasp} 

The main reason for the lack of applications of hybrid functionals to extended systems is the considerable computational cost of evaluating the exact exchange energy, 
particularly within the plane wave-pseudopotential 
approach that is most frequently used for electronic structure calculations. 
This has limited most applications to systems with a small unit cell. When large supercells are needed, 
like e.g. in {\it ab initio} molecular dynamics (AIMD) simulations\cite{CPMD}, a screened exchange approximation \cite{HSE}
is offen used to alleviate the computational burden of hybrid functionals\cite{PBE0_water}. 

In this work, we present an accurate and efficient scheme to compute the exact exchange energy and potential for large molecules and extended insulating systems. Our scheme can be easily included in existing plane wave codes and has computational cost that scales linearly with system size. The approach is based on a unitary transformation of the occupied subspace from Bloch to (maximally) localized Wannier functions (MLWFs)\cite{MLWF}. MLWFs are exponentially localized and, since the exchange between two orbitals is restricted to the spatial region of orbital overlap, the amplitude of the exchange interaction between two MLWFs decays rapidly with the distance between their centers. 
Thus, typically each Wannier orbital exchanges only with a finite number of neighboring orbitals and the number of pair interactions per orbital is independent of system size \cite{MLWF_review}. As a result, our procedure to compute exact exchange is order-N, i.e. its computational cost scales linearly with system size. We demonstrate the effectiveness of our approach in two representative applications using the PBE0 \cite{PBE0} hybrid functional for exchange and correlation. In one we perform a ground state electronic and structural 
optimization for crystalline silicon, in the other we perform a finite temperature AIMD simulation for the same system.

In the following we assume, for simplicity, a closed-shell system with $N/2$ doubly occupied one-electron states. Extension to spin-polarized systems is straightforward.
The PBE0 \cite{PBE0} total energy functional can be written as:
\begin{eqnarray}
E^{\rm PBE0}& = &-\frac{1}{2}\sum_i \langle \varphi_i \vert
\nabla^2 \vert \varphi_i \rangle + \int V_{\rm ion}( {\bf r})n( {\bf r}) d{\bf r}\cr
&+& \frac{1}{2}\int \int \frac{n({\bf r})n({\bf r'})}{\vert {\bf r}-{\bf r'} \vert}
d {\bf r}d{\bf r'} 
+  E_{\rm ion}[\{ {\bf R_I} \} ] \cr
&+& E_{xc}^{\rm PBE0}  ,
\label{eq:PBE0_total_energy}
\end{eqnarray}
where $n({\bf r})=2\sum_{i=1}^{N/2} \vert \varphi_i({\bf r}) \vert^2$
is the electronic density, 
$N$ is the total number of electrons, and the $\varphi_i$ are the occupied one-electron orbitals,
and atomic unit (a.u.: $\hbar=m=e^2 =1$) are adopted. 
As in standard DFT formulations using LDA or GGA functionals, 
the first four terms in Eq.~(\ref{eq:PBE0_total_energy}) represent the electronic kinetic energy, the potential energy of the electrons in the field of the nuclei, the average electrostatic interaction among the electrons and the electrostatic repulsion between the nuclei, respectively. Here we adopt a pseudopotential formulation. Thus the sums extend to the valence states only while $n({\bf r})$ and $\varphi_i$ denote pseudo-density and pseudo-wavefunctions, respectively .
The last term on the right hand side of Eq.~(\ref{eq:PBE0_total_energy}) 
is the PBE0 exchange correlation energy \cite{PBE0}, $E_{xc}^{\rm PBE0}$ given by: 
\begin{equation}
E_{xc}^{\rm PBE0} = \frac{1}{4}E_x+\frac{3}{4}E_{x}^{\rm PBE} + E_{c}^{\rm PBE}.
\label{eq:PBE0_energy}
\end{equation}
Here $E_x$ denotes exact exchange, $E_x^{\rm PBE}$ is the PBE exchange, 
and $E_c^{\rm PBE}$ is the PBE correlation functional \cite{PBE}.
The exact exchange energy $E_x$ has the Hartree-Fock expression in terms of the one-electron (pseudo-)orbitals:
\begin{equation}
E_x = -{2} \sum_{i,j}
\int\int \frac{\varphi_{i}^{\ast}({\bf r}) \varphi_{j}^{\ast}({\bf r'})
\varphi_{j}({\bf r}) \varphi_{i}({\bf r'}) } {\vert {\bf r}- {\bf r'} \vert } 
 d {\bf r'} d {\bf r} \, ,
\label{eq:Exx_energy}
\end{equation}

The ground state energy is obtained by minimizing the energy 
functional, Eq.~(\ref{eq:PBE0_total_energy}), with respect to the occupied orbitals.
This leads to the one-particle equations:
\begin{eqnarray}
 \Bigl [ -\frac{1}{2}\nabla^2 &+& V_{\rm ion}({\bf r}) + V_{\rm H}({\bf r}) 
 + \frac{3}{4}V_x^{\rm PBE}({\bf r})
+ V_c^{\rm PBE}({\bf r}) 
 \Bigr]\varphi_{i}({\bf r}) \cr
&+& \frac{1}{4}\int V_x({\bf r, r'}) \varphi_{i}({\bf r'})d{\bf r'}
= \varepsilon_i \varphi_{i}({\bf r}) ,
\label{eq:PBE0_potential}
\end{eqnarray}
where
$V_{\rm H}({\bf r})$ and $V_{\rm ion}({\bf r})$ are the Hartree and the ionic (pseudo-)potentials, respectively. $V_x^{\rm PBE}({\bf r})$ and $V_c^{\rm PBE}({\bf r})$, the PBE exchange and correlation potentials, depend on the electron density and its gradient at position $\bf r$. The exact exchange potential ${V}_x({\bf r, r'})$ is the non-local integral operator of Hartree-Fock theory. It is given by:
\begin{equation}
{V}_x({\bf r, r'})= -2 \sum_j 
\frac{\varphi_j^*({\bf r'})\varphi_j({\bf r})}{\vert {\bf r} - {\bf r'} \vert }
\label{eq:Exx_potential}
\end{equation}
We notice that the above procedure is not strictly a Kohn-Sham scheme. The latter would require an exchange potential given by the functional derivative of the exchange energy with respect to the electron density rather than with respect to the orbitals. Since the explicit functional dependence of the exact exchange energy on the density is not known, implementation of a strict Kohn-Sham scheme would require a special procedure such as e.g. the Optimized Effective Potential (OEP) method \cite{Kummel}. The latter would be considerably more computationally expensive than our approach while giving 
essentially the same ground state energies \cite{Kummel}. 

The action of $\hat{V}_x({\bf r, r'})$ on the orbital $\varphi_i$ in 
Eq.~(\ref{eq:PBE0_potential}) is an orbital dependent term $D_x^i ({\bf r} )$ given by:
\begin{equation}
D_x^i ({\bf r} ) \equiv \frac{\delta E_x}
{\delta \varphi_i^*} = -2  \sum_j \int d {\bf r'}
\frac{ \varphi_j ^* ( {\bf r'}) \varphi_i ({\bf r'}) \varphi_j ( {\bf r} ) }
{ \vert {\bf r} - {\bf r'} \vert }
\label{eq:action}
\end{equation}
Eq.~(\ref{eq:action}) shows that $D_x^i ({\bf r} )$  includes the exchange interactions of the orbital $\varphi_i$ with all the occupied orbitals $\varphi_j$ (including the self-interaction). 
Usually in extended system implementations \cite{PBE0_vasp}, 
each pair interaction in Eq.~(\ref{eq:action}) is evaluated in reciprocal 
space taking advantage of the convolution theorem \cite{footnote}
\begin{equation}
 \int d {\bf r'}
\frac{ \varphi_j ^* ( {\bf r'}) \varphi_i ({\bf r'})  }
{ \vert {\bf r} - {\bf r'} \vert } \; \rightarrow 4\pi 
\frac{\rho_{ij}({\bf G})}{\vert {\bf G} \vert^2} \; ,
\label{eq:FFT}
\end{equation}
where  $\rho_{ij}({\bf G})$ is the Fourier Transform of $\rho_{ij}({\bf r})=\varphi_i({\bf r})\varphi_j({\bf r})$. This can be calculated using the Fast Fourier Transform 
(FFT) algorithm at a cost proportional to $N_{\rm FFT}ln(N_{\rm FFT})$, where $N_{\rm FFT}$ is the size of the plane wave grid. Thus, if the functions $\{\varphi_i\}$ are delocalized throughout the entire supercell, evaluating   Eq.~(\ref{eq:FFT}) for all orbital pairs would result in an overall computational effort proportional to $N^2\times N_{\rm FFT}ln(N_{\rm FFT})$. Neglecting the weak logarithmic dependence, this amounts to cubic scaling with size. While plane-wave LDA or GGA calculations have cubic scaling with size, they only require a number of FFTs that scales linearly with $N$. The need to perform a number of FFTs that scales quadratically with $N$ is what makes traditional plane wave implementations of the hybrid functional method very expensive.

Instead of evaluating the exact exchange in terms of delocalized 
Bloch orbitals $\{\varphi_i\}$, we choose to work with MLWFs $\{\widetilde\varphi_i\}$. This requires a unitary transformation of the occupied subspace, $ \widetilde\varphi_i = \sum_{j=1}^{N/2} U_{ij} \varphi_j $, which leaves the ground state energy invariant. In terms of the MLWFs $D_x^i ({\bf r})$ becomes:
\begin{equation}
D_x^i ({\bf r} )  =  -2 \bigl( \; v_{ii}({\bf r}) \widetilde\varphi_{i}({\bf r})
+ \sum_{i \neq j}v_{ij}({\bf r}) \widetilde\varphi_{j}({\bf r}) \; \bigr )
\label{eq:orbital_dependent}
\end{equation}
where the self-interaction ($v_{ii}$) and the pair-exchange ($v_{ij}$) 
potentials satisfy the Poisson equations: 
\begin{equation}
\nabla^2 v_{ii}  =  - 4\pi\widetilde\rho_{ii} \; , \; 
\nabla^2 v_{ij}  =  - 4\pi\widetilde\rho_{ij} 
\label{eq:Exx_Wannier}
\end{equation}
Here, $\widetilde\rho_{ij}({\bf r})=\widetilde\varphi_i({\bf r})\widetilde\varphi_j^*({\bf r})$.
In the hybrid functional formalism the contribution associated to $v_{ii}({\bf r})$ in 
Eq.~(\ref{eq:orbital_dependent}) partially cancels the spurious self-interaction present 
in the Hartree potential $V_{\rm H}({\bf r})$ in Eq.~(\ref{eq:PBE0_potential}).
The contribution associated to the pair potential $v_{ij}({\bf r})$ in 
Eq.~(\ref{eq:orbital_dependent}) gives the exchange 
interaction for two electrons of equal spin residing in different orbitals. The potential $v_{ij}$ 
(when $i$ is either equal to or 
different from $j$) can be viewed as the electrostatic potential generated by the charge 
distribution $\widetilde\rho_{ij}({\bf r})$.

\begin{figure}[ht]
\includegraphics[width=2.2in]{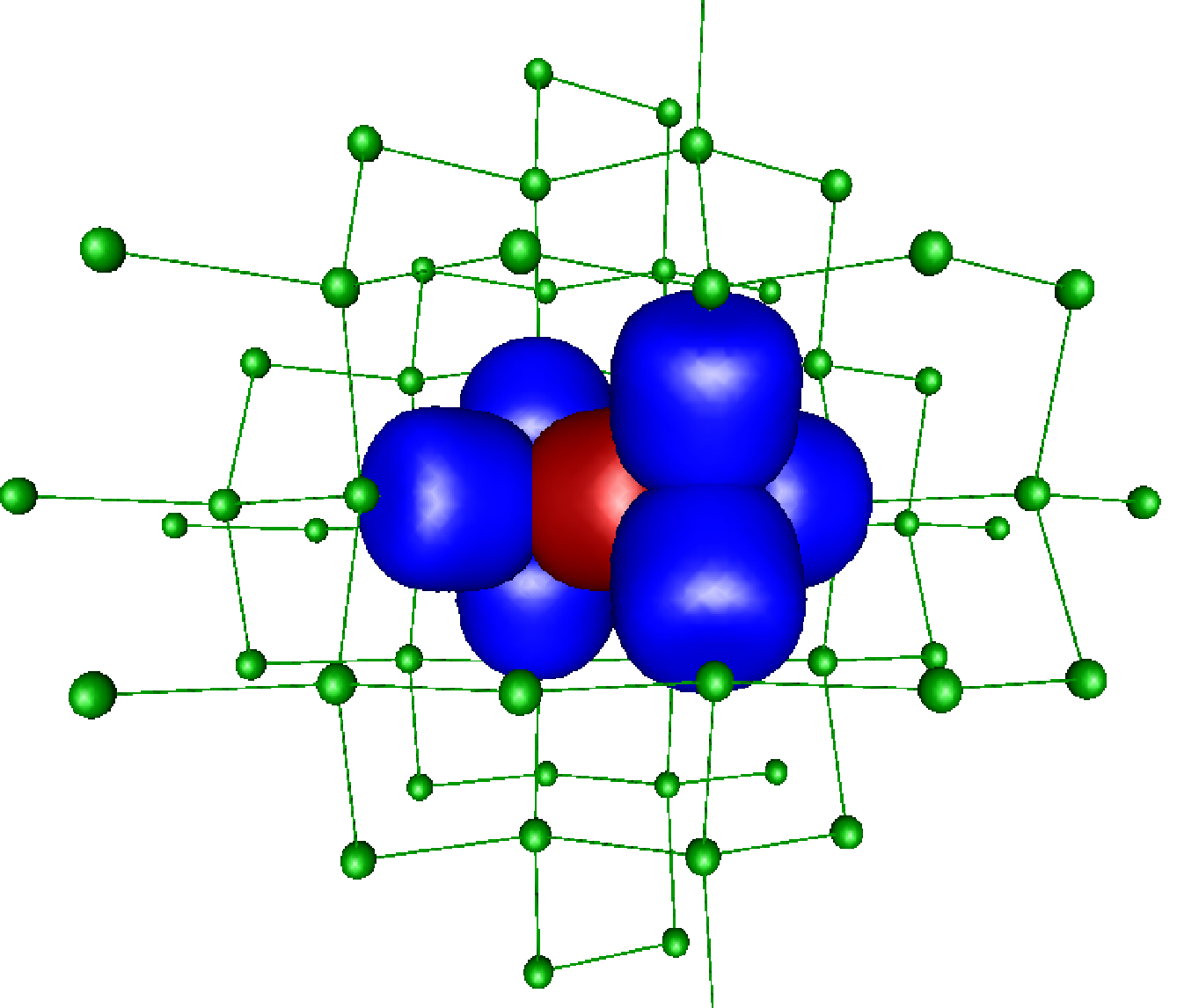}
\caption{\label{fig1} (Color online.) Overlap between a tagged Wannier orbital
(red) and its nearest neighboring Wannier orbitals (blue) in the 
64-atom Si supercell. Si atoms are denoted by the green spheres. }
\end{figure}

In the Wannier representation it is convenient to work in real space. This point is illustrated 
in Fig.~\ref{fig1}. Since the exchange interaction is only present in 
the region where two orbitals overlap, i.e. where $\widetilde\rho_{ij}\neq 0$, 
the pair potential $v_{ij}({\bf r})$ is conveniently calculated by solving the 
corresponding Eq.~(\ref{eq:Exx_Wannier}) in a spatial region  
significantly smaller than the simulation cell. Moreover only a small subset 
of orbitals $\widetilde\varphi_j$ contribute to the exchange interaction with a 
tagged orbital $\widetilde\varphi_i$.  

We have implemented the above method in the CP code of the Quantum-Espresso package \cite{QuantumEspresso}. 
In the following, we apply our approach to compute the electronic ground state, to optimize the cell parameter, 
and to carry out an AIMD simulation for crystalline Si in the diamond structure using the PBE0 functional. 
In these calculations we used supercells ranging from 64 to 216 atoms. In all the calculations we used a PBE norm-conserving pseudopotential with ($3s3p$) valence. The plane-wave energy cutoff was 15 Ry and we sampled the Brillouin Zone at the 
$k=0$ point ($\Gamma$ point). For comparison we also performed PBE0 calculations with the same pseudopotential and plane-wave cutoff using the conventional reciprocal space method to calculate exact exchange as implemented in the PWSCF code of Quantum-Espresso. These calculations were performed on the Si 2-atom unit cell using a large set of $k$ points to sample the Brillouin Zone.

In our approach, we first perform a ground state calculation using the semi-local PBE functional. 
Then, given the PBE Kohn-Sham eigenstates $\{\varphi_i({\bf G})\}$ we construct the corresponding 
MLWFs $\{\widetilde\varphi_i({\bf G})\}$ (in reciprocal space) by iteratively minimizing the spread 
functional \cite{Manu}. The corresponding MLWFs in real space, $\{\widetilde\varphi_i({\bf r})\}$, 
are obtained by FFT and are represented on a uniform real space mesh. In the Si diamond structure, 
each MLWF is centered in the midpoint between two adjacent atoms, and overlaps significantly 
with the six nearest neighboring orbitals, as shown in Fig.~\ref{fig1}.

Since the density $\widetilde\rho_{ij}({\bf r})$ is known for each pair of orbitals, 
we can associate with each pair of orbitals an orthorhombic box with 
sides $(l_x^c, l_y^c, l_z^z)$ such that outside this box $\widetilde\rho_{ij}({\bf r})$ is 
smaller than a given cut-off value $\rho^{\rm cut}$, which we take 
equal to $2 \times 10^{-4}\ {\rm bohr}^{-3}$ in the present work. 
We then solve Eq.~(\ref{eq:Exx_Wannier}) inside the box. 
Notice that the box contains a greatly reduced set of grid points compared to the simulation cell. 
For example, in our 64-atom Si calculation the real space grid needed to compute the pair 
potential $v_{ij}$ generated by two adjacent orbitals contains only ~ 20\% of the mesh points of the simulation cell. 
Even fewer points are needed to compute the pair potential $v_{ij}({\bf r})$ generated by more distant 
orbitals. Since the density $\widetilde\rho_{ij}({\bf r})$ is vanishingly small when the distance 
between the orbitals $i$ and $j$ is sufficiently large, many pair interactions are negligibly small. 
We find that in our 64-atom Si supercell each orbital exchanges appreciably only 
with 30 orbitals out of the set of 127 neighboring orbitals.   

To solve the Poisson equation the Laplace operator $\nabla^2$ is discretized on 7 mesh points. 
The resulting finite difference equation has the form of a linear matrix equation 
of the type ${\bf A}{\bf x}={\bf b}$. The symmetric and positive definite square
 matrix ${\bf A}$ is sparse and has dimension $n$, 
where $n$ is the number of mesh points inside the reduced box. 
The vector ${\bf x}$ corresponds to the unknown $v_{ij} ({\bf r})$, 
and the (known) vector $b$ corresponds to the pair density $\widetilde\rho_{ij}({\bf r})$. 
The values of $v_{ij}({\bf r})$ at the boundary of the box are set by the multipole expansion:
\begin{equation}
v_{ij}({\bf r})=4\pi \sum_{l,m}\frac{1}{2l+1}
  q_{lm} \frac{ Y_{lm}(\theta, \phi)}{{ r}^{l+1}} 
\label{eq:multipole_expansion}
\end{equation}
where the multipoles $q_{lm}$ are given by the integrals:
\begin{equation}
q_{lm}=\int Y_{lm}^*(\theta ', \phi '){ r'}^l\widetilde\rho_{ij}({\bf r'})d{\bf r'} \; .
\label{eq:multipoles}
\end{equation}
In Eq.~(\ref{eq:multipoles}) the $Y_{lm}$ are spherical harmonics referred to the center of the pair density, 
which we define by 
${\bf R}_{ij}^c \equiv {\int {\bf r} \widetilde \rho_{ij}({\bf r})}/{\int \widetilde \rho_{ij}({\bf r})}$. 
We found that inclusion of multipoles up to $l=6$ is sufficient to achieve good accuracy.

Solving the linearized Poisson equation ${\bf A}{\bf x}={\bf b}$ is equivalent to finding the vector ${\bf x}$ that minimizes the function 
$f({\bf x}) = \frac{1}{2}{\bf x}^T{\bf A}{\bf x}-{\bf b}^T{\bf x} + c$, where $c$ is an arbitrary constant. This minimization is efficiently performed with the conjugate gradient (CG) method \cite{CG}. We terminate the CG iteration when the residue in the calculation of $v_{ij}({\bf r})$ is everywhere smaller than 10$^{-5}$ a.u.. In order to calculate the $D_x^i$ in Eq.~(\ref{eq:orbital_dependent}) we need to evaluate the products $v_{ij}({\bf r})\widetilde\varphi_j({\bf r})$ in the region where 
$\vert \widetilde \varphi_j({\bf r})\vert ^2 > \rho^{cut}$. 
This region may include points outside the box associated to 
the pair density $\widetilde \rho_{ij}({\bf r})$ but values of $v_{ij}({\bf r})$ outside that box are easily obtained from the multipole expansion in  Eq.~(\ref{eq:multipole_expansion}). 

Having calculated the $D_x^i({\bf r})$, the PBE0 ground state is obtained by conventional electronic structure methods. Here we optimize the electronic degrees of freedom via damped second order Car-Parrinello dynamics\cite{CPMD, Tassone} in which the "force" acting on the orbitals, $H^{\rm PBE0}\widetilde\varphi_i({\bf r})$, includes the additional $D_x^i({\bf r})$ terms to account for exact exchange. Finally, the exchange energy $E_{x}$ is given by the sum of the energies of the orbital pairs in presence of the corresponding pair potential $v_{ij}({\bf r})$,
\begin{equation}
E_{x} = {-2}\sum_{ij}\int \widetilde\varphi_i({\bf r}) \widetilde\varphi_j({\bf r})v_{ij}({\bf r})d{\bf r} \, .
\label{eq:exchange_decomp}
\end{equation}
The exchange energy in equation (Eq.~(\ref{eq:exchange_decomp})) can be viewed as a sum of orbital
contributions $e_x(i): E_x = \sum_ie_x(i)$. The $i-th$ orbital contribution $e_x(i)$ can be 
further decomposed into self-exchange $e^{\rm self}(i) = \int \widetilde\varphi_i^2 v_{ii} $ and pair-exchange
$e^{\rm pair}(i) = \sum_{j \neq i} \int \widetilde\varphi_i \widetilde\varphi_j v_{ij}$.
\begin{table}[ht]
\caption{Contributions to the exchange energy $e_x$ (in a.u.) from shells of neighbors.
 $R(I)$ is shell radius (bohr) and $N(I)$ is the coordination number of shell $I$. 
The experimental lattice constant $a_0=5.43 \AA$ is used.}
\begin{ruledtabular}
\begin{tabular}{lddddd}
Shell $I$ & \multicolumn{1}{c}{\quad 0} 
& \multicolumn{1}{c}{\quad 1} 
& \multicolumn{1}{c}{\quad 2} &
\multicolumn{1}{c}{\quad 3} 
& \multicolumn{1}{c}{\quad 4} \\
\hline
$R(I)$ & 0 & 3.63 & 6.28 & 7.26 & 8.11  \\ 
$N(I)$ & \multicolumn{1}{c}{1} & \multicolumn{1}{c}{6} & 
\multicolumn{1}{c}{12} & \multicolumn{1}{c}{12} & \multicolumn{1}{c}{12} \\
$ e_x(I) $ & -0.465 & -0.059 & -0.002 & -0.007 & -0.0001 \\ 
\end{tabular}
\end{ruledtabular}
\label{table:Pair}
\end{table}

In Table \ref{table:Pair} we report the calculated exchange energy per orbital, 
in crystalline Si using a 64-atom supercell. In this system the MLWFs are all 
equivalent by symmetry, i.e the orbital index in $e_x(i)$ can be dropped. 
Moreover the MLWF centers coincide with the bond centers and it is convenient 
to group the pair exchange contributions into contributions originating from the 
different shells of neighbors of a bond center. The Table lists the shell index $I$ 
(which is 0 for the central site, 1 for the first shell of neighbors, etc.), 
the corresponding shell radius $R(I)$, the corresponding coordination number $N(I)$, 
and the corresponding exchange energy contribution $e_x (I)$, with $e_x = \sum_I e_x(I)$.

It is evident that the largest contribution to $e_x$ comes from the self-interaction $e_{x}(0)$, 
and that the exchange contributions of the neighboring shells, $e_{x}(I)$, with $I = 1,2,..$, 
goes rapidly to zero with increasing shell radius. As a matter of fact the exchange energy 
contribution of the fourth shell is only $1/300-th$ of the contribution due the first shell of neighbors. 

\begin{table}[ht]
\caption{Comparison of our real space method and the reciprocal space method implemented in PWSCF.
$E$ denotes total (pseudo-) energy per atom (Rd) and VBW is the valence band witdth (eV).}
\begin{ruledtabular}
\begin{tabular}{ldddd}
 & \multicolumn{2}{c}{\quad Our approach} 
& \multicolumn{2}{c}{\quad PWSCF }
  \\
\hline
${\bf k}-{\rm points}$ & \multicolumn{2}{c}{\rm  Gamma} & \multicolumn{1}{c}{\quad $4\times 4 \times 4$ }
& \multicolumn{1}{c}{\quad $6 \times 6 \times 6$ }\\ 
$N_{\rm atom}/{\rm cell}$ & \multicolumn{1}{c}{\quad 64} & \multicolumn{1}{c}{\quad 216} 
& \multicolumn{1}{c}{\quad 2} & \multicolumn{1}{c}{\quad 2}  \\ 
$E$ & -7.865 & -7.870 &  -7.867 & -7.873  \\
VBW & 13.3 & 13.3 & 13.3 & 13.3  \\
\end{tabular}
\end{ruledtabular}
\label{table:PBE0_energy}
\end{table}
In table \ref{table:PBE0_energy}, we report the calculated PBE0 ground-state energy using two supercells, one with 64 atoms and one with 216 atoms. The results of the two calculations are compared to the results obtained with the conventional reciprocal space method using a 2-atom unit cell. In the case of the two large supercells we used $\Gamma$ point sampling, while we used two large sets of $k$ points in the conventional calculations as indicated in the Table. The two sets of calculations are in very close agreement: the valence band widths are the same, while the slight differences in total energy can be attributed to the differences in the $k$-point sampling. 
\begin{table}[ht]
\caption{Lattice constant $a_0$(bohr) and bulk modulus $B_0$(GPa) of a Si. }
\begin{ruledtabular}
\begin{tabular}{lddd}
 & \multicolumn{1}{c}{\quad Our approach} 
& \multicolumn{1}{c}{\quad PWSCF } 
& \multicolumn{1}{c}{\quad Expt.\tablenotemark[1]} \\
\hline
$a_0$ & 5.49 & 5.49 
& 5.43  \\ 
$B_0$ & \multicolumn{1}{c}{\quad 100} & \multicolumn{1}{c}{\quad 99} &  
 \multicolumn{1}{c}{\quad 99}  \\
\end{tabular}
\end{ruledtabular}
\label{table:lattice_constant}
\tablenotetext[1] {Ref.~\protect\onlinecite{Si}.}
\end{table}

As a further comparison we report in Table III the equilibrium lattice constant $a_0$ 
and the bulk modulus $B_0$ calculated with a 64-atom Si supercell. 
We also report in the same table the results of a conventional calculation with 
a 2-atom unit cell and a 4x4x4 $k$-point grid. Again, the results of the two calculations are in excellent agreement.

\begin{figure}[ht]
\includegraphics[width=3.0in]{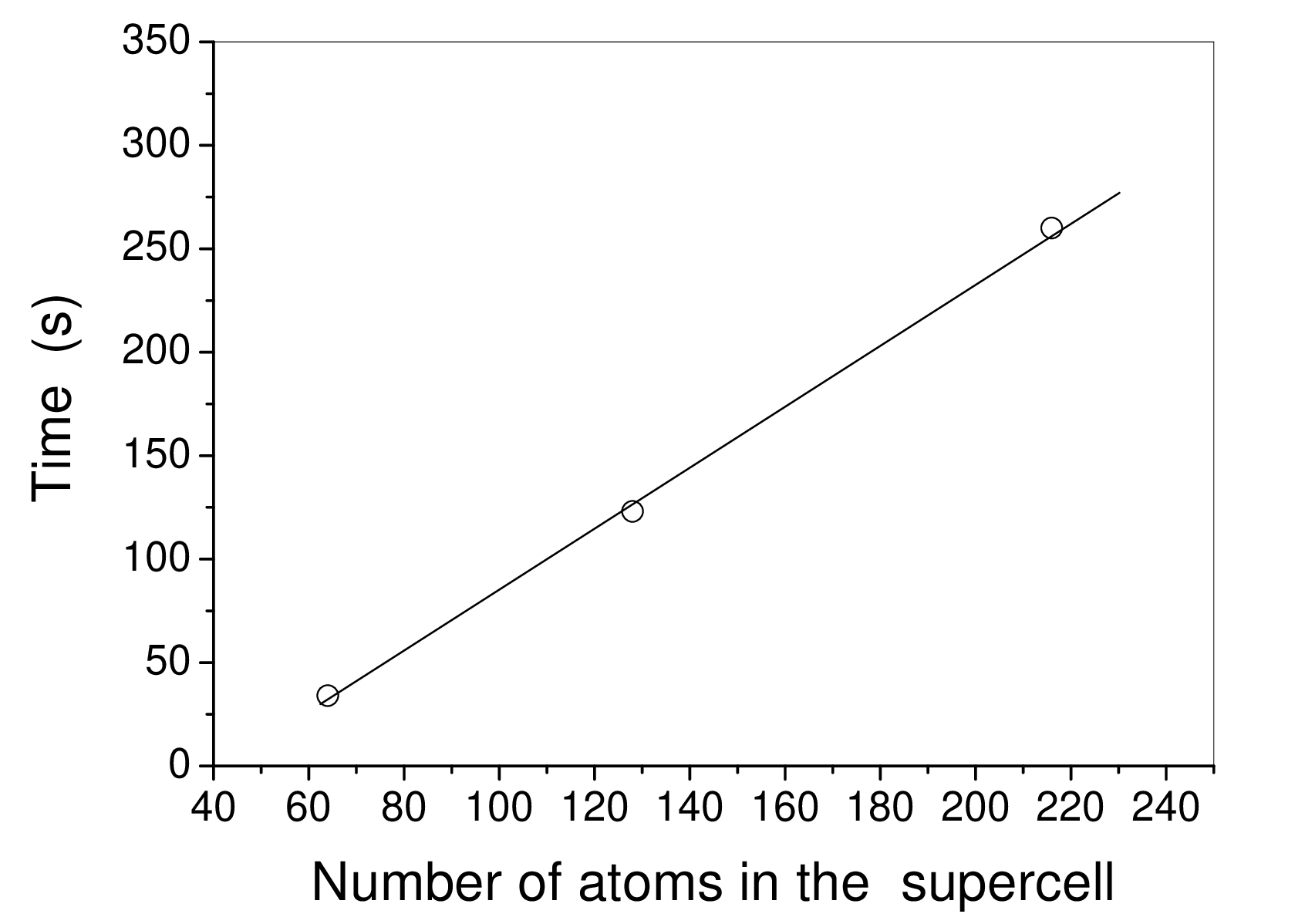}
\caption{\label{fig2} Computational cost of exact exchange per iteration of AIMD dynamics with different supercells.
The computational cost is given by the time (s) necessary to compute exact exchange on a
16-CPU 3.2 Ghz Intel Xeon computer cluster. }
\end{figure}

In our approach, the computational cost of an exact exchange calculation depends on the number of pair exchanges that need to be included to achieve a desired accuracy. Since each orbital has exchange only with a finite number of neighboring orbitals 
independently of the system size, the computational effort of the exact exchange calculation should scale linearly with system size. Fig.~\ref{fig2} shows that this is indeed the case.  

Finally, we demonstrate that our approach makes AIMD simulations with hybrid functionals, such as PBE0, feasible at a modest computational cost. In AIMD simulations a large number of time steps, typically tens of thousands, are necessary to obtain statistically meaningful results. As a consequence AIMD simulations with hybrid functionals are very challenging and so far have only been performed by making some approximation, like the screened exchange approximation, in the calculation of the exchange integrals \cite{PBE0_water}. In our approach we do not need to modify the Coulomb potential to eliminate exchange interactions at large distance. These are automatically truncated by the exponential decay of the MLWFs and all the relevant pair exchange interactions are included. To show the feasibility of AIMD simulations we tested our approach in a finite temperature simulation of a Si sample with 64 atoms in a simple cubic supercell geometry. The simulation was initiated by randomly displacing the atoms from their crystalline sites while their velocities were set to zero. The subsequent trajectories were obtained by numerically integrating the Car-Parrinello equations of motion with the standard Verlet algorithm \cite{RC review}. MLWF-based AIMD trajectories were generated as described in Ref. \onlinecite{Manu}, using the PBE0 total energy functional $E^{\rm PBE0}$ to compute the forces on electronic and ionic degrees of freedom. 
\begin{figure}[ht]
\includegraphics[width=3.2in]{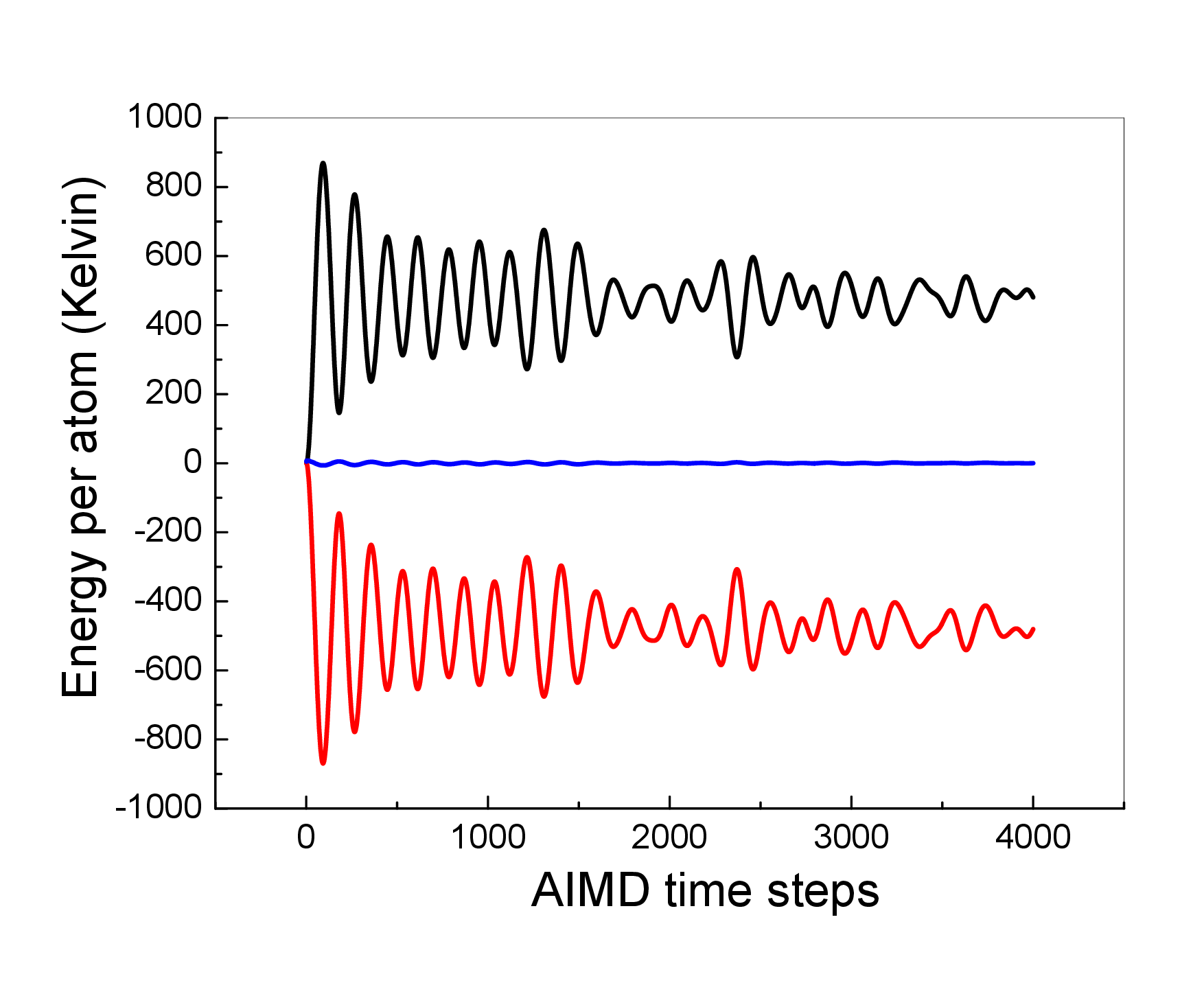}
\caption{\label{fig3} (Color online.) Potential energy per atom $E^{\rm PBE0}$  (red line),
kinetic energy per atom $K$ (black line) and internal energy per atom $U$ (blue line)
vs. AIMD time steps. The average $U$ sets the zero of the energy scale. A time step $\Delta$ = 8 a.u. 
and a fictitious electronic mass of 800 a.u. were used.}
\end{figure}

We plot in fig.~\ref{fig3} the time variation along a nuclear trajectory of $E^{\rm PBE0}$, i.e the potential energy of the ions (nuclei plus core electrons), of their kinetic energy $K = 1/2\sum_IM_I\dot{\rm R}^2_I$, and of the ionic internal energy $U = K + E^{\rm PBE0}$. The internal energy is an exact constant of motion of classical nuclear dynamics but is only approximately constant in Car-Parrinello simulations due to the fictitious dynamics of the electrons. Fig.~\ref{fig3} shows that indeed $U$ is approximately constant with minor fluctuations and no drift over the time scale of the simulation. This is the typical behavior observed in standard simulations of insulating systems based on LDA or GGA functionals. We conclude that our real space treatment of exact exchange does not lead to any appreciable degradation of the quality of the integrated trajectories compared to standard AIMD simulations. 

The AIMD trajectory reported in fig.~\ref{fig3} was obtained on a 16 CPUs PC cluster and took 34 s of real time per time step. For comparison a standard GGA simulation for the same system would take only 2.5 s per time step on the same computational platform. This example shows that while hybrid functional calculations remain more expensive than GGA calculations, AIMD trajectories lasting for many ps are possible with access to moderate computer resources. Moreover the order-N cost of the exact exchange calculation means that the overhead of hybrid functional calculations should be a comparatively smaller fraction of the overall computational cost in simulations on bigger systems.      

In conclusion we have developed an order-N method to compute exact exchange in extended insulating systems. By exploring the locality of maximally 
localized Wannier functions, we calculate the orbital dependent exchange potential and the corresponding exchange energy contribution directly in real space. The approach is sufficiently efficient to make AIMD simulations with hybrid functionals possible and can be effectively implemented on parallel computer platforms. Its computational efficiency should be even better for large band-gap systems such as e.g. water, where the MLWFs are more localized than in silicon. Since exact exchange is a basic ingredient in many-body approaches to electronic excitations, such as e.g. the GW scheme \cite{Hedin}, our approach should facilitate the application of these schemes to systems requiring large supercells, such as liquids and disordered systems in general \cite{Wei}.

\acknowledgments
We would like to thank Morrel H Cohen, Eric Walter and Andrew Rappe for useful discussions.
This work has been supported by the Department Of Energy under grant
DE-FG02-06ER-46344, grant DE-FG02-05ER46201 and by AFOSR-MURI  F49620-03-1-0330. 



\begin{thebibliography}{0}
\bibitem{DFT}
See e.g. R. G. Parr and W. Yang, {\it Density Functional Theory of Atoms and Molecules} (Oxford
University Press, New York, 1989).

\bibitem{Yang}
A. J. Cohen, P. Mori-S\,anchez, and W. Yang, Science {\bf 321}, 792 (2008).

\bibitem{Hybrid}
A. D. Becke, J. Chem. Phys. {\bf 98}, 1372 (1993).

\bibitem{Kummel}
S. K\"{u}mmel, L. Kronik, Rev. Mod. Phys. {\bf 80}, 3 (2008).

\bibitem{Hybrid_solid_chemistry}
F. Cor\`{a}, M. Alfredsson, G. Mallia, D. S. Middlemiss, W. C. Mackrodt, 
R. Dovesi, and R. Orlando, Struct. Bonding (Berlin) {\bf 113}, 171 (2004).

\bibitem{localized_state}
C. di Valentin, G. Pacchioni, and A. Selloni, Phys. Rev. Lett. {\bf 79}, 1905 (2006).

\bibitem{PBE0_vasp}
M. Marsman, J. Paier, A. Stroppa, and G. Kresse, J. Phys.: Condens. Matter {\bf 20}, 
064201 (2008).

\bibitem{CPMD}
R. Car and M. Parrinello, Phys. Rev. Lett. {\bf 55}, 2471 (1985).

\bibitem{HSE}
J. Heyd, G. E. Scuseria, J. Chem. Phys. {\bf 118}, 8207 (2003).

\bibitem{PBE0_water}
T. Todorova, A. Seitsonen, J. Hutter, I-Feng. Kuo and C. Mundy, 
J. Chem. Phys. B {\bf 110}, 3685 (2006).

\bibitem{MLWF}
N. Marzari and D. Vanderbilt, Phys. Rev. B {\bf 56}, 12847 (1997).

\bibitem{MLWF_review}
N. Marzari, I. Souza, and D. Vanderbilt, Highlight of the Month, 
Psi-K Newsletter {\bf 57}, 129 (2003).

\bibitem{PBE0}
J. P. Perdew, Ernzerhof, and K. Burke, J. Chem. Phys. {\bf 105}, 9982 (1996).

\bibitem{PBE}
J. P. Perdew,K. Burke, and M. Ernzerhof, Phys. Rev. Lett. {\bf 77}, 3865 (1996).

\bibitem{footnote}
We assume here that the periodic supercell is sufficiently 
large that the sampling of the Brillouin Zone can be limited to the $k=0$ point only. 

\bibitem{QuantumEspresso}
See http://www.quantum-espresso.org and http://www.pwscf.org. 

\bibitem{Manu}
M. Sharma, Y. Wu, and C. Car, Int. J. Quant. Chem. {\bf 95}, 821 (2003).

\bibitem{CG}
W. H. Press, S. A. Teukolsky, W. T. Vetterling, and B. P. Flannery, {\it Numerical Recipes} 
(Cambridge University Press, Cambridge, 1992).

\bibitem{Tassone}
F. Tassone, F. Mauri, and R. Car, Phys. Rev. B {\bf 50}, 10561 (1994).

\bibitem{Si}
J. Heyd, G. E. Scuseria, J. Chem. Phys. {\bf 121}, 1187 (2004).

\bibitem{RC review}
R. Car, in {\it Conceptual Foundations of Materials: 
A Standard Model for Ground- and Excited-State Properties, 
Contemporary Concepts of Condensed Matter Science},
 edited by S. G. Louie and M. L. Cohen (Elsevier, Amsterdam, 2006), Chap. 3, p. 64.

\bibitem{Hedin}
L. Hedin, Phys. Rev. {\bf 139}, A796 (1965).

\bibitem{Wei}
W. Chen, X. Wu and R. Car, in preparation. 

\end{thebibliography}
\end{document}